# High Quality Queueing Information from Accelerated Active Network Tomography

Tommaso Rizzo
Enrico Fermi Center
Rome, Italy
tommaso.rizzo@inwind.it

József Stéger
Eotvos University
Budapest, Hungary
steger@complex.elte.hu

Péter Pollner
Stat. and Biol. Phys. Res. Group of HAS
Budapest, Hungary
pollner@complex.elte.hu

István Csabai
Eotvos University
Budapest, Hungary
csabai@complex.elte.hu

Gábor Vattay
Eotvos University
Budapest, Hungary
vattay@complex.elte.hu

## ABSTRACT

Monitoring network state can be crucial in Future Internet infrastructures. Passive monitoring of all the routers is expensive and prohibitive. Storing, accessing and sharing the data is a technological challenge among networks with conflicting economic interests. Active monitoring methods can be attractive alternatives as they are free from most of these issues. Here we demonstrate that it is possible to improve the active network tomography methodology to such extent that the quality of the extracted link or router level delay is comparable to the passively measurable information. We show that the temporal precision of the measurements and the performance of the data analysis should be simultaneously improved to achieve this goal. In this paper we not only introduce a new efficient message-passing based algorithm but we also show that it is applicable for data collected by the ETOMIC high precision active measurement infrastructure. The measurements are conducted in the GEANT2 high speed academic network connecting the sites, which is an ideal test ground for such Future Internet applications.

## 1. INTRODUCTION

The possibility of obtaining fast and reliable network performance estimates is a vital step in order to perform ambitious tasks such as the design of refined and more efficient traffic-control and dynamic-routing protocols, and the detection of anomalous and/or malicious behavior. Measurement, collection, and analysis of network data such as link delay and loss is unfeasible on large scale since subnetworks might have conflicting economic interests. While there is no gain for individual routers to collect performance statistics, it costs considerably in terms of computation, hardware and maintenance. Furthermore the transmission of this information to some central control node would create considerable networking overhead. Network tomography[17] is an emerging field of research whose goal is performance estimation of decentralized, unregulated and heterogeneous networks such as the Internet. The basic idea of Network tomography is to perform active measurements that do not require special coordination from the network operators and cause limited traffic overhead. These measurements are related to the unobserved quantities of interest such as delays inside the network. Statistical inference techniques are applied in order to recover this hidden information. The term *Network Tomography* was introduced by Vardi[17] to stress the similarity between inferential problems in network and medical tomography. As it relies on statistics collected only at the end hosts its infrastructure is scalable and can offer lower hardware and maintainance costs.

In order to make it a viable alternative of direct passive monitoring we should overcome some existing problems: i) The network tomography computation requires a summation over all possible combination of internal delays in the network and has a non-polynomial computational complexity. ii) The temporal resolution is low since the processed primary network measurements do not have sufficiently high precision. Also the computational complexity limits the resolution achievable in realistic runtime. iii) The computation does not scale well with the size of the network.

In this paper we show that it is possible to handle these problems and to realize large scale network tomography in practice. Our most important result is that using a new modified version of the classical message-passing algorithm the computations of network tomography can be accelerated substantially. It is shown that the computation of the likeli-hood function and its numerical maximization through the Expectation-Maximization (EM) algorithm can be done recursively, thus we can introduce a highly reliable estimator through a fast computation of polynomial complexity. The linear scaling of the runtime of the new algorithm with the network size is shown numerically.

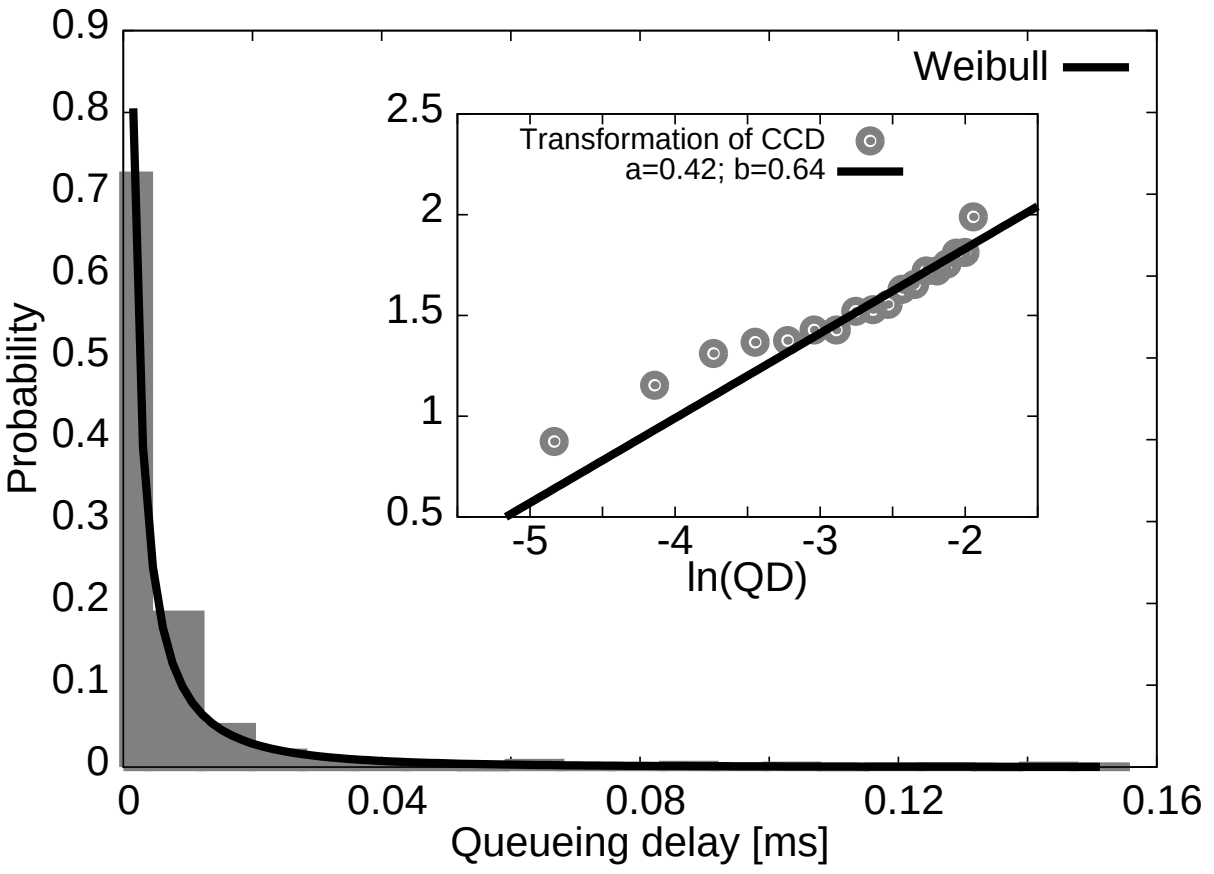

**Figure 1: An example of queueing delay distribution for a single up-link queue determined by our new iterative algorithm from high temporal resolution delay data. Inset The logarithm of the complementary cumulative distribution of the Weibull distribution $P(X > x) = \exp(-bx^a)$ is a power-law function. The logarithm of the measured complementary cumulative distribution is shown on a log-log scale to derive the parameters of the Weibull approximation. The slope of the fitted line yields $a$, which is related to the Hurst-exponent $a = 2(1 - H)$ and $H = 0.79$.**

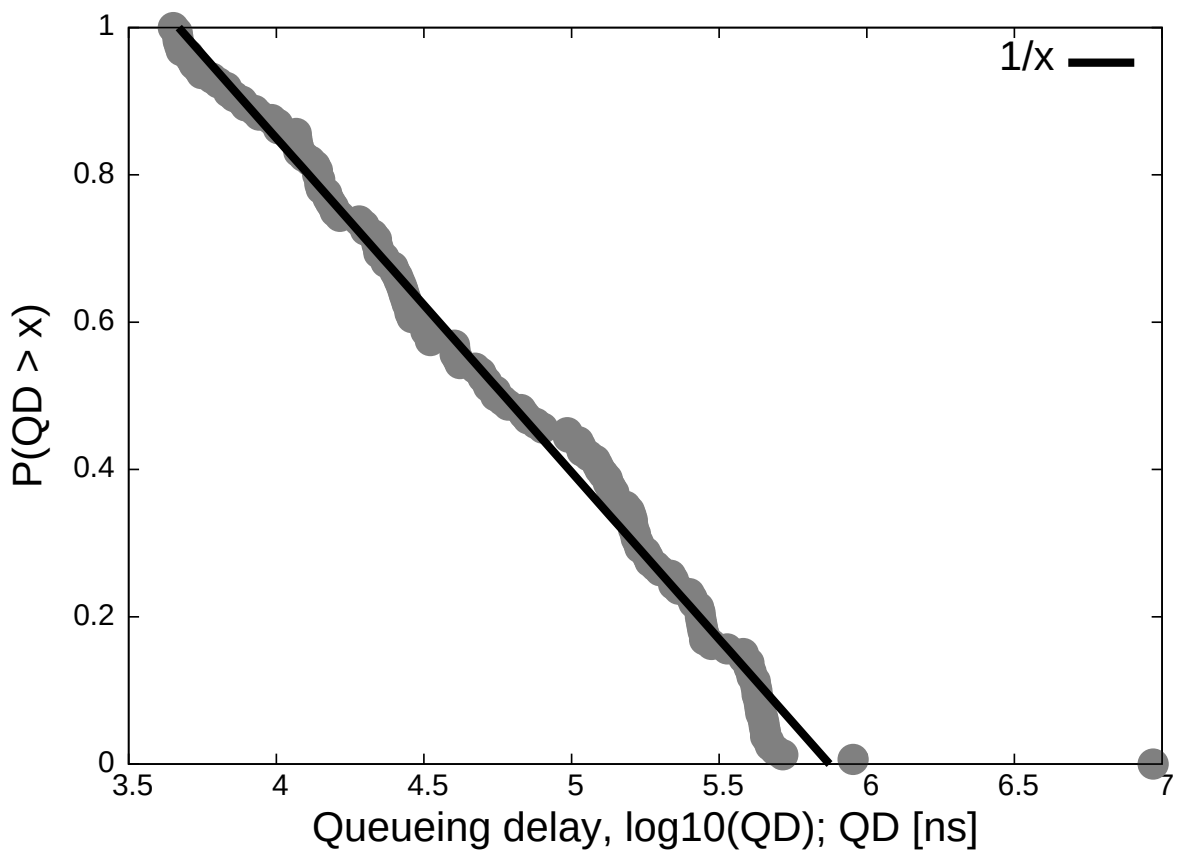

**Figure 2: A new scaling law for queuing delays in the Internet. The complementary cumulative distribution of the average queueing delays on a semi-logarithmic plot. The continuous line is a $\sim -\log(x)$ distribution corresponding to a $\sim 1/x$ type scaling of the corresponding probability density of average delays.**

This accelerated method can then be used for processing high temporal resolution measurements. For testing our methods on real data we conduct measurements in the ETOMIC active probing infrastructure[4], where special DAG 3.6GE measurement cards and GPS time synchronization of the measurement hosts provide one-way delay data with 100 ns absolute precision.

The combined result of the fast algorithm and the high temporal resolution is that we can infer internal delay distributions with high resolution. This new richness of detail makes it possible to analyze the shape of these distributions further. We find cases when our tomography topology permits to resolve single hop links. We analyze the tail distribution of a single queue (see Fig. 1) and even the Hurst exponent of the self-similar input traffic on the link can be determined. Another result demonstrated in our experiment is that hundreds of delay distributions can be measured simultaneously by using about ten measurement sites. The timescales of distributions span three orders of magnitude and a new network wide spatial scaling law of average queuing delays is discovered (see Fig. 2).

The plan of the paper is as follows: in Sec. 2 we introduce the model and framework and we discuss previous results for the ML estimator and the EM algorithm. In Sec. 3 we introduce our new recursive algorithm and provide some basic numerical tests on it. In Sec. 4 we introduce the measurement infrastructure, topology discovery and one-way delay measurements. In Sec. 5 we present the issues of the numerical implementation of the inference algorithm. In Sec. 6 we show newly discovered scaling properties of the inferred queuing delay data.

## 2. QUEUEING DELAY DISTRIBUTION INFERENCE

In this section we review the problem of queueing delay distribution inference. Consider a tree as the one depicted in Fig. 3. Each node is labeled by a number $i$ and link $i$ connects node $i$ it to its parent node. During an observation a batch of back-to-back packets are sent from the root to all the leaves. At branching points packets are directed towards their destination leaf. The process continues until the packets reach the leaves. At the leaves we measure the end-to-end delays the probe packets experience in the network. From this information the delay distributions of the internal links are to be inferred. The overall delay observed at a receiver is given by the sum of the delays $x_i$ on all the links where the packet has travelled. Thus for a given batch of packets the directly observable end-to-end delays $y_j$ ($j$ running over all leaves) are related to the internal link delays $x_i$ by the following linear combination

$$Y = AX, \tag{1}$$

where $A$ is the routing matrix. For instance given a scenario of a four-leaf binary tree of Fig. 3 the relation reads:

$$\begin{pmatrix} y_4 \\ y_5 \\ y_6 \\ y_7 \end{pmatrix} = \begin{pmatrix} 1 & 1 & 0 & 1 & 0 & 0 & 0 \\ 1 & 1 & 0 & 0 & 1 & 0 & 0 \\ 1 & 0 & 1 & 0 & 0 & 1 & 0 \\ 1 & 0 & 1 & 0 & 0 & 0 & 1 \end{pmatrix} \begin{pmatrix} x_1 \\ x_2 \\ \vdots \\ x_7 \end{pmatrix}. \tag{2}$$

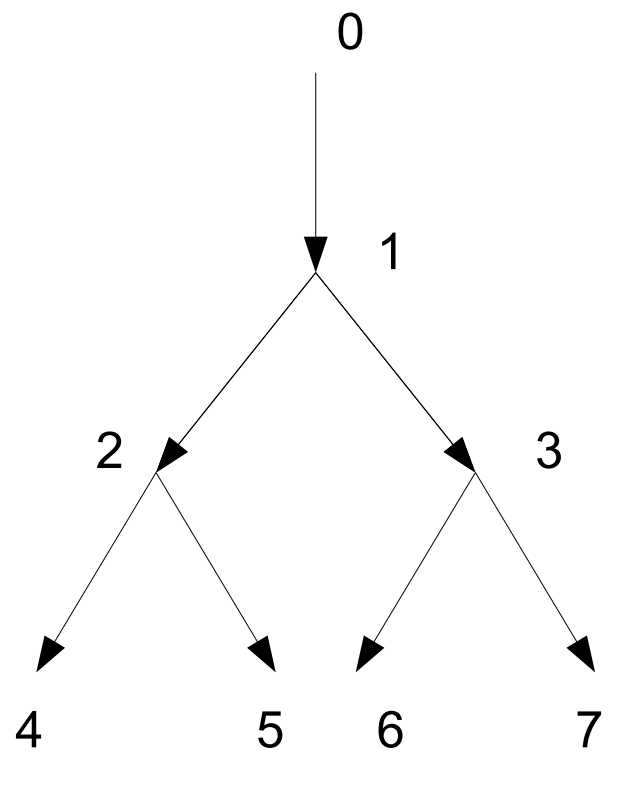

**Figure 3: A four-leaf binary tree**

In our inference model we use the following assumptions:

- Observed end-to-end delays are independent random variables for each observation time $t$.
- During the experiment the routing matrix $A$, determined by the network topology and the routing tables, is constant for all observations.
- Each hidden link delay $x_i$ experienced in a given measurement is independent and identically distributed random variable characterized by a delay distribution $\theta_i(x_i)$. We denote the collective set of these parameters as $\Theta$.

We adopt the notation $Y^t = (y_j)$ , $t = 1, \dots T$, for all observed delays at the leaves and $X^t$ for the unobserved internal-link delays in a given observation $t$; $\{Y^t\}$ and $\{X^t\}$ represent the whole set of $T$ observations. For a given set of link parameters $\Theta$ the likelihood of different observations is factorized:

$$P(\{Y^t\}|\Theta) = \prod_{t=1}^{T} P(Y^t|\Theta) \tag{3}$$

The likelihood of a given observation at time $t$ is given by:

$$P(Y^t|\Theta) = \sum_{X^t} P(X^t|\Theta) 1_{Y^t = AX^t} \tag{4}$$

where $1_{Y^t=AX^t}$ is the indicator function on the set of possible values of $X^t$ such that they satisfy the routing equation $Y^t = AX^t$. Its computation requires the summation over all possible values of the link-delays $X^t$ leading in general to an exponential complexity.

In [3, 16] it is claimed that internal delay probabilities $\Theta$ can be uniquely related to the joint probabilities $\gamma$ of observing a certain set of delays at the leaves.[1] The resulting set of equations $\Theta = \Gamma^{-1}(\gamma)$ can be computed iteratively, the procedure involving the solution of symbolic polynomial equations. The corresponding estimator $\hat{\Theta}$ was defined estimating the probabilities of the events at the leaves with their observed frequencies $\hat{\gamma}$ and plugging these estimates in the exact equations. The fact that $\hat{\gamma}$ is a consistent and asymptotically normal estimator of $\gamma$ ensures that $\hat{\Theta}$ is also consistent and asymptotically normal. The advantage of this method is that it is very fast because the equations are obtained recursively. At the same time estimating the probabilities $\gamma$ with the frequencies $\hat{\gamma}$ seems reasonable, at least when the number of observation is very large.

[1] This identifiability property is true for the so-called *canonical trees*, defined as those that satisfy the condition $\theta_i(0) > 0$ $\forall i$. This condition maintains that none of the loss probabilities for the links is 1.

In [2] authors introduce Maximum-likelihood estimators to infer loss probabilities for the internal segments of a network tree and in [3, 16] the approach was generalized to link-delay inference. These are obtained through the well-known formula:

$$\Theta_{ML} = \text{argmax}_{\Theta} \ln p(Y|\Theta) \tag{5}$$

where $\ln p(Y|\theta)$ is the log-likelihood, *i.e.* the logarithm of the probability of observing $Y$ given that the parameters of the network are $\Theta$. The Maximum-Likelihood method is a central and well-established method of inference. Standard theorems ensure that the ML estimator is consistent and asymptotically normal and make it the standard for parameter estimation.

In a subsequent paper [12] the iterative estimates were compared with Maximum-Likelihood estimates. The two methods yield the same result in the case of loss inference, but in the case of delay estimation it was shown in [12] that ML produces different estimates from those of [2]. The difference between the two approaches can be understood considering as parameters to be estimated the set of probabilities $\gamma$ instead of the set of probabilities $\Theta$. Since there is a bi-unique relationship $\Theta = \Gamma^{-1}(\gamma)$ between them (which can be computed recursively through the algorithm of [16]) the ML estimate of $\gamma$ is given by $\gamma_{ML} = \Gamma(\Theta_{ML})$. Although in some cases $\Gamma_{ML}$ may be equal [2] to the observed frequencies $\hat{\Gamma}$ they are different in general. Furthermore it was shown in [12] that ML yields more reliable estimates than those of the iterative method. Given that both estimators are consistent, the effect disappears when the number of observations goes to infinity, but can be rather dramatic for a finite number of observations, see Fig. 5 of [12]. In [12] it is suggested that the effect originates from the fact that the mapping $\theta = \Gamma^{-1}(\gamma)$ is obtained solving symbolic polynomial equations whose roots can be unstable.

## 2.1 The ML estimator and the EM algorithm

The main problem in obtaining the ML estimator is the maximization of the likelihood. Candidates are the extrema but in general it is not possible to find analytical solutions for the extremization equations. Following Ref. [12] we use the Expectation-Maximization algorithm, a well-known method for missing-value problems which is guaranteed to converge to a local maximum of the likelihood function. Given a log-likelihood $\ln \sum_X P(X, Y|\Theta)$ to be extremized, the EM algorithm yields a series of converging estimates $\Theta^n$ through the iterative formula:

$$\Theta^{n+1} = \text{argmax}_{\theta} \sum_X P(X|Y, \Theta^n) \ln P(Y, X|\Theta) \tag{6}$$

This can be seen as a two-step process, in the first step (Expectation) the expectation of the function $\ln P(Y, X|\Theta)$ is computed with respect to the probability $P(X|Y, \Theta^n)$, in the second step (Maximization) the expectation is maximized with respect to $\Theta$. The basic result concerning EM is that during the EM iteration step the value of the true likelihood

[2] The Gaussian distribution is a classical example in which estimating the average of the distribution with the sample average is equivalent to estimating it through ML.

function is non-decreasing, in particular if it has a unique maximum point it will converge to it.

In the present context the function to be maximized in the EM algorithm is given by:

$$Q(\Theta, \Theta^n) = \sum_{\{X^t\}} P(\{X^t\}|\{Y^t\}, \Theta^n) \ln P(\{Y^t\}, \{X^t\}|\Theta) \tag{7}$$

The expectation step requires the knowledge of the function $P(\{X^t\}|\{Y^t\}, \Theta)$, which is factorized over different observations:

$$P(\{X^t\}|\{Y^t\}, \Theta) = \prod_{t=1}^{T} P(X^t|Y^t, \Theta). \tag{8}$$

The distribution of the delays on a given observation is also factorized, provided they satisfy the routing equation:

$$P(X^t|Y^t, \Theta) = C_t \prod_i P(x_i^t|\theta_i) 1_{Y^t = AX^t} \tag{9}$$

where $C_t$ is a observation-dependent normalization constant. On the other hand the function $P(\{Y^t\}, \{X^t\}|\Theta)$ can be also written in a factorized form:

$$P(\{Y^t\}, \{X^t\}|\Theta) = \prod_{t=1}^{T} \left[ \prod_i P(x_i^t|\theta_i) 1_{Y^t = AX^t} \right] \tag{10}$$

When computing the expectations, after some manipulations we get:

$$Q(\Theta, \Theta^n) = \sum_{t=1}^{T} \sum_{i=1}^{N} \sum_{X^t} P(X^t|Y^t, \theta^n) \ln P(x_i^t|\theta_i) \tag{11}$$

where $N$ is the total number of nodes and $P(x_i^t|\theta_i)$ is the probability distribution of the delay on link $i$. Following [16] we parameterize it through a normalized function $\theta_i(x)$ such that:

$$P(x_i^t|\theta_i) = \theta_i(x_i^t) \tag{12}$$

where the possible values of $x_i$ includes $x_i = \infty$, that corresponds to the loss of the packet. Introducing Lagrange multipliers to ensure the normalization of the $\theta_i(x)$ the bound function Eq. (11) can be extremized differentiating explicitly with respect to $\theta_i(x)$. The resulting equations lead to:

$$\theta_i^{n+1}(x) = C_i \sum_{t=1}^{T} \frac{P(Y^t, x_i^t = x|\Theta^n)}{P(Y^t|\Theta^n)} \tag{13}$$

where the $C_i$ is an $x$-independent proportionality constant to be determined through the normalization condition. In the above equation $P(Y^t|\Theta^n)$ is the probability that the final delays are $Y^t$ when the link parameter is $\Theta^n$ and $P(Y^t, x_i^t = x|\Theta^n)$ is the probability that the final delays are $Y^t$ and that the delay at node $i$ is $x$ when the link parameter is $\Theta^n$.

On the other hand the computation of the Likelihood requires a summation over all possible internal delays and in principle has an exponential computational complexity. In order to avoid such a computation, Yu and Liang (2003) introduced a pseudo-likelihood estimator, obtained maximizing not the true likelihood but a more tractable approximation of it. They show that, much as ML and the recursive estimator, the pseudo-likelihood estimator is also consistent and asymptotically normal.

Next we show that the computation of the likelihood and its numerical maximization through the EM algorithm can be done in a recursive way rigorously without the introduction of approximations and pseudo-likelihood estimators. Thus we have a highly reliable estimator through a fast computation of polynomial complexity.

## 3. RECURSIVE COMPUTATION OF THE LIKELIHOOD

Although in principle the computation of the two quantities needed to perform the EM iteration step in Eq. (13) has exponential complexity, they can be computed recursively on the tree in a message-passing fashion[5]. In the following we redefine the notation in order to simplify the presentation. According to Eq. (13) each observation can be treated separately so in the following we omit the index $t$. Given a node $i$, $f(i)$ represents its father on the tree, that is the node that sends the packets to it, while $d(i)$ represents the set of nodes that are connected to $i$ through a single link, that is the set of nodes that receive their packets directly from $i$. The delay $x_i$ is defined as the difference between the time when the packet leaves node $f(i)$ (the father of $i$ on the tree) and the time when node $i$ sends the messages to its descendants $d(i)$. The delay depends on the properties of the physical link $i - j$ and on buffering delays at node $i$. We define $y_i$ as the set of delays observed at the leaf nodes *connected* to node $i$, *i.e.* those that are the leaves of the sub-tree whose root is node $i$. Accordingly $y_0$ is the whole set of observed delays ($Y$ in the notation of the previous section) since each node at the leaves is connected to the root 0. The quantity $R_i$ is defined as the total delay of the packet when it is sent by node $i$. In order to implement the EM algorithm, according to Eq. (13), we need two quantities: $P(y_0|\Theta)$ and $P(y_0, x_i|\Theta)$. $P(y_0|\Theta)$ is the total probability of observing the delays $y_0$ while $P(y_0, x_i|\Theta)$ is the probability of observing the delays $y_0$ and a delay $x_i$ on node $i$.

All the probabilities we are considering in this section are constrained to a given fixed value of $\Theta$, therefore in the following, for simplicity, we omit also the argument $\Theta$. We present the approach in the case of continuous values of $x$ and $R$ including the value $\infty$ corresponding to the loss of the packet, the generalization to discretized distributions being straightforward. In order to compute $P(y_0)$ we introduce for each node the quantities $P(y_j|R_j)$ that is the probability of observing delays $y_j$ at the leaves connected to $j$ given that $j$ sent its messages with a delay $R_j$. These quantities obey a recursive equation:

$$\begin{aligned} P(y_j|R_j) &= \prod_{i \in d(j)} \int dR_i dx_i P(x_i) P(y_i|R_i) \times \\ &\times \delta(R_i - (R_j + x_i)) \end{aligned} \tag{14}$$

where the factor $\delta(R_i - (R_j + x_i))$ enforces the condition $R_i = R_{f(i)} + x_i$, *i.e.* it is a Dirac delta function when $R_i$, $R_j$ or $x_i$ are all finite while it take values 0 or 1 when either $R_i$, $R_j$ or $x_i$ is infinite. The functions $P(y_j|R_j)$ can be computed recursively starting from the leaves of the tree where the following condition holds:

$$P(y_i|R_i) = \delta(R_i - y_i) \quad \text{for } i \in \text{leaves} \tag{15}$$

Then the total probability $P(y_0)$ can be computed as $P(y_0|0)$.[3]

[3]Note that in practice it is useless to consider the case in which more than one node is directly connected to the root,

In order to compute the quantity $P(y_0, x_i)$ we introduce a new set of functions $P(R_j, y_{0/j})$. This is the joint probability that there is a delay $R_j$ at node $j$ and that $y_{0/j}$ delays are observed at all the leaves but those connected to $j$.

Given a node $i$ whose father is $j = f(i)$, $P(R_i, y_{0/i})$ obeys a recursive equation:

$$\begin{aligned} P(R_i, y_{0/i}) &= \int dx_i dR_j P(x_i) P(R_j, y_{0/i}) \times \\ &\times \ \delta(R_i - (R_j + x_i)) && (16) \\ P(R_j, y_{0/i}) &= P(R_j, y_{0/j}) P(y_{j/i}|R_j) && (17) \\ P(y_{j/i}|R_j) &= \prod_{k \in d(j)/i} \int dx_k dR_k P(x_k) P(y_k|R_k) \times \\ &\times \ \delta(R_k - (R_j + x_k)) && (18) \end{aligned}$$

Using the set of functions $P(y_i|R_i)$ previously computed, these equations can be solved recursively going down on the tree from the root where the following condition holds:

$$P(R_0, y_{0/0}) = \delta(R_0) \qquad (19)$$

The knowledge of the functions $P(R_i, y_{0/i})$ allows to compute the probability $P(y_0, x_i)$ which is given by:

$$\begin{aligned} P(y_0, x_i) &= \int dR_j dR_i P(R_j, y_{0/i}) P(x_i) P(y_i|R_i) \times \\ &\times \ \delta(R_i - (x_i + R_j)) && (20) \end{aligned}$$

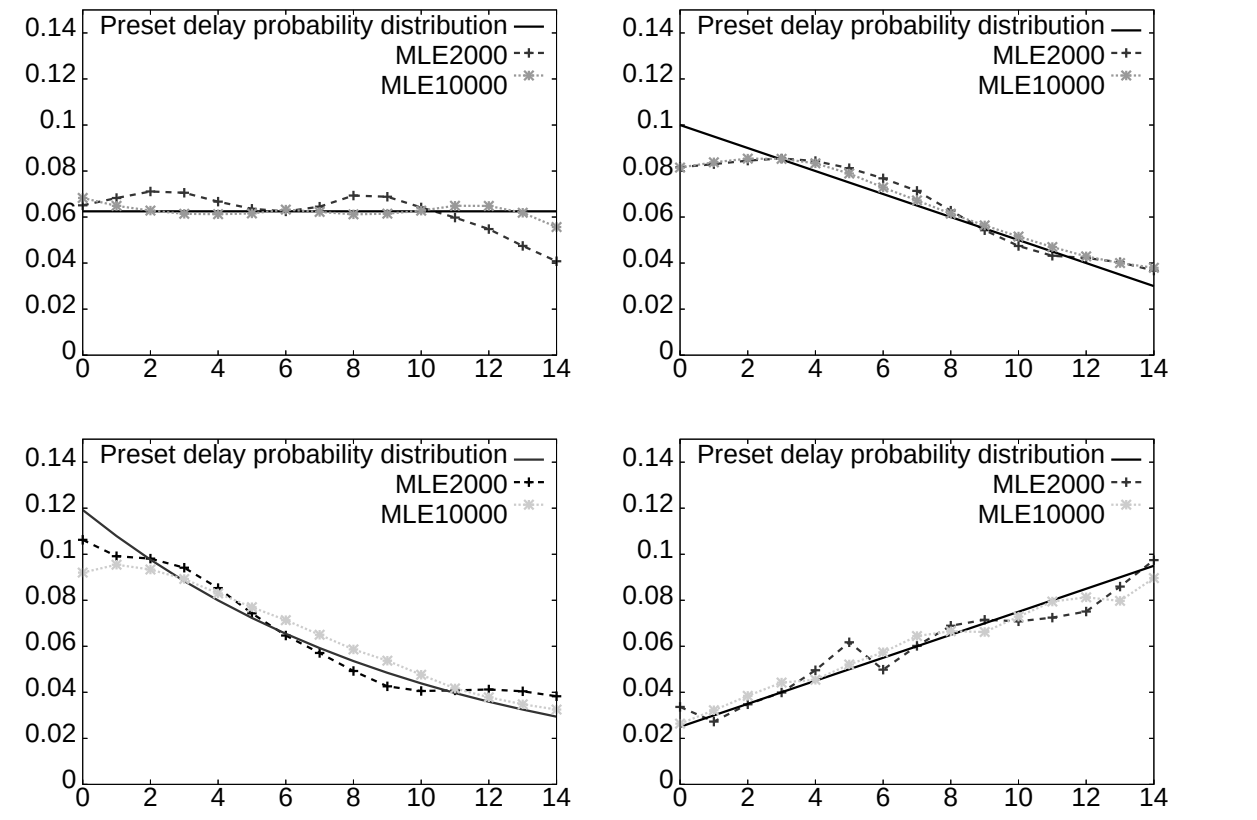

**Figure 4: True and ML inferred link delay distribution of links 1, 4, 13 and 85 from 2000 and 10000 measurements, on an 85-nodes tree (see text)**

We have applied the present algorithm to the analysis of a simulated experiment. In order to make contact with previous results, the parameters of the model are those of Ref. [12]. In particular the range of possible delays is discretized with values going from 0 to 14 plus the value $\infty$. We generate $T = 2000$ and $T = 10000$ iid multicast measurements and make inference on the internal delay distributions through the EM algorithm. As observed by Yu and Liang (2003) the $T = 2000$ is the $T$-regime where the recursive algorithm of Ref. [16] could sometime yield imprecise results. We consider a multicast tree with the topology of a regular tree with 3-levels and bifurcation number $c = 4$, accordingly the total number of nodes is 85 and the total number of parameter to be estimated is $85 \times 16 = 1360$. Note that by means of the recursive ML algorithm we obtain ML estimates for system sizes ten times bigger than those reachable through the exponential computation of ML. In subfigures in Fig.4 we report the true and inferred ML values for some links at different levels of the tree.

because they would correspond to completely uncorrelated trees

# 4. LARGE-SCALE QUEUEING DELAY TOMOGRAPHY MEASUREMENT

Multicast tomography was tested in the MINC project[6]. A similar approach using unicast end-to-end measurements is also used in Ref. [3]. In this section we show the measurement infrastructure we used to conduct queueing delay tomography experiments on a regular basis in the European Internet. We highlight the most important implementation issues and show how to resolve them to collect reliable data for the analysis. We briefly sketch the steps of the tomography experiment and give the configuration parameter set. Then we review the main issues of the computation process.

## 4.1 The measurement infrastructure

The European Traffic Observatory Measurement InfrastruCture (ETOMIC) is a modern high-precision, synchronized measurement platform. It consists of 18 measurement nodes, the infrastructure (see Fig. 5) provides the ability to perform a rich variety of experiments with different probing techniques. Measurement nodes are operating in the 100 Mbps ethernet LANs of various academic institutes in Europe(see [7] for more details). Each institute is served by the high speed European academic research network GEANT and the hosts are typically 2-4 hops away from the high speed core. An ETOMIC measurement node is basically a standard PC, hosting an additional interface card (Endace DAG 3.6GE) designed for precise active and passive measurements. They embed a separate high stability reference clock that provides very accurate time-stamping of the IP probe packets, with a time-resolution of 60 ns. The pulse-per-second reference signal comes from GPS devices connected directly to the DAG cards, thus the clocks are globally synchronized on the 100 ns scale.

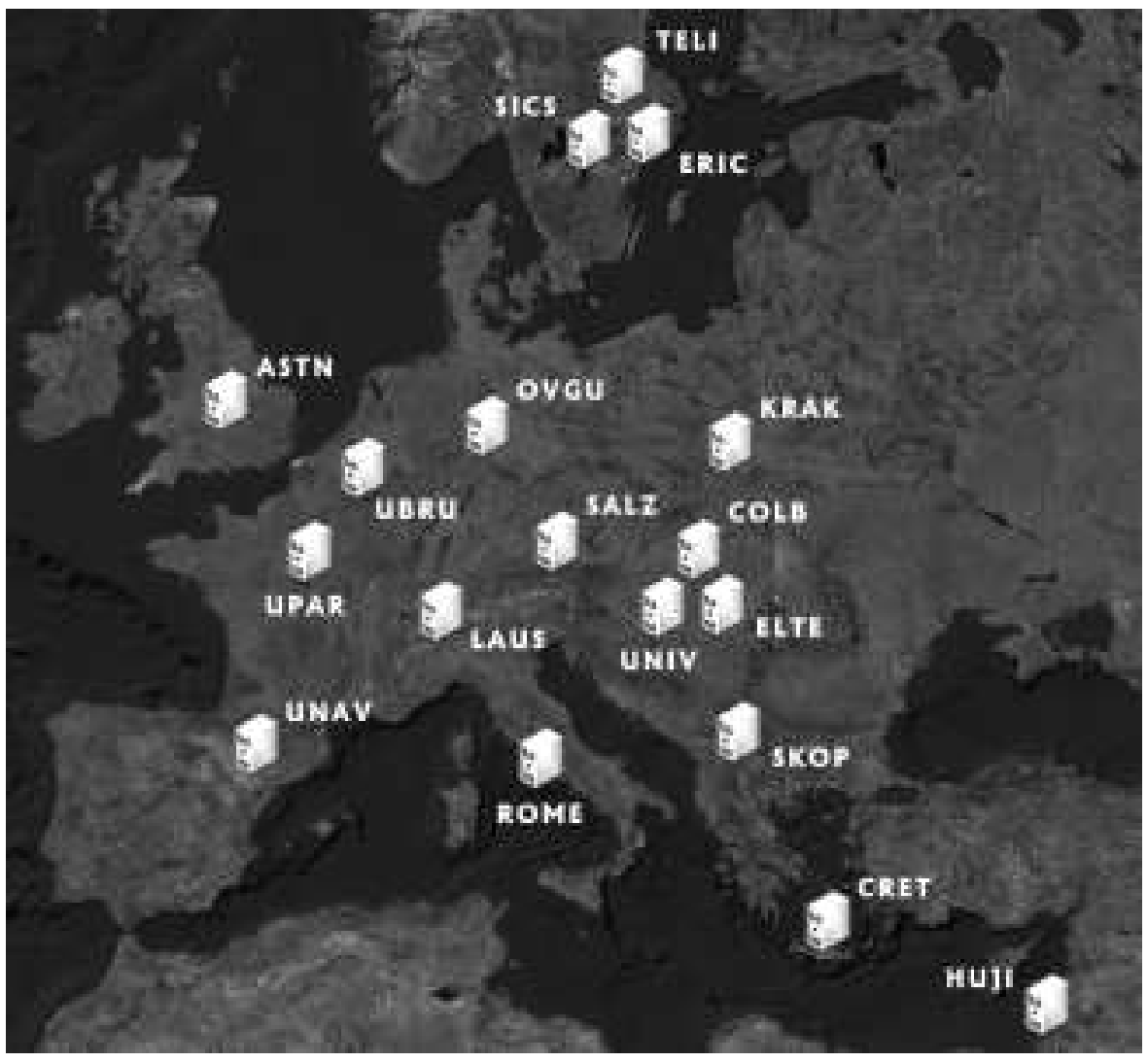

**Figure 5: The measurement stations of the ETOMIC infrastructure. Measurement nodes are operating at various academic institutes in Europe.**

## 4.2 Topology discovery

We started the discovery of network topology connecting the hosts of ETOMIC with standard `traceroute`. Several load balancing routers were discovered which make the network paths ambiguous. We then changed the procedure and modified the `traceroute` program to use UDP probe packets with fixed IP address and port number pairs. This way we managed to fix the network path as the load balancing routers discovered in our experiment seem to route these packets in a unique manner. This is much in the flavor of the recently introduced Paris Traceroute[1]. As a first attempt, we can construct the tree graph from the sequences of IP addresses produced by the modified `traceroute`. IP addresses are considered as nodes and they are connected if they are successive hops in a path. By using the long term network path observations in the ETOMIC infrastructure, the missing IP addresses showing up as $\star$ in the `traceroute` output can also be identified or at least distinguished from each other. There are certain cases when the resulting graph does not span a proper branching tree and is not directly suitable for our purposes. A branching tree comprises one root (the node where packets are sent from), several leaves (the receiver nodes) and inner nodes. The root node should have only one child, otherwise the tree can be split up into independent trees starting with the children of the root node. The leaves have no children by definition. The inner nodes are branching nodes if they have two or more children. From a general tree-like graph one can construct a branching tree by replacing sequences of nodes with single child with a compound link connecting the branching nodes (or leaves in the end) directly. Network tomography will resolve the delay distributions of these compound links, which might consist of several hops in the network. In certain cases in our experiment it is possible to find and resolve single hop links. In general the number of hops in compound links decreases with increasing number of observation points.

A more important issue is that the topology of the traceroute graph is not always a tree. This is due to the fact that routing in real networks not always follows the shortest network path principle. There exist rear cases when a path that splits off at some branching point joins another segment of the graph again. Nodes at the joining points have more than one fathers (see Fig. 6), which is not allowed in a tree. As the probe packets visit these segments at different time instances – which can be carefully checked in the experiments – they do not interact with each other and pose no real problem later at the ML evaluation step. We can handle the situation simply by introducing clones of the nodes and handling them as if they were separate nodes in each affected network path. We connect the clones to their new fathers. If there are more than one branching and merging points along two paths, all common nodes after the first branching point (counted from the root) have to be handled as separate logical nodes.

## 4.3 Measuring the one-way delays of probes

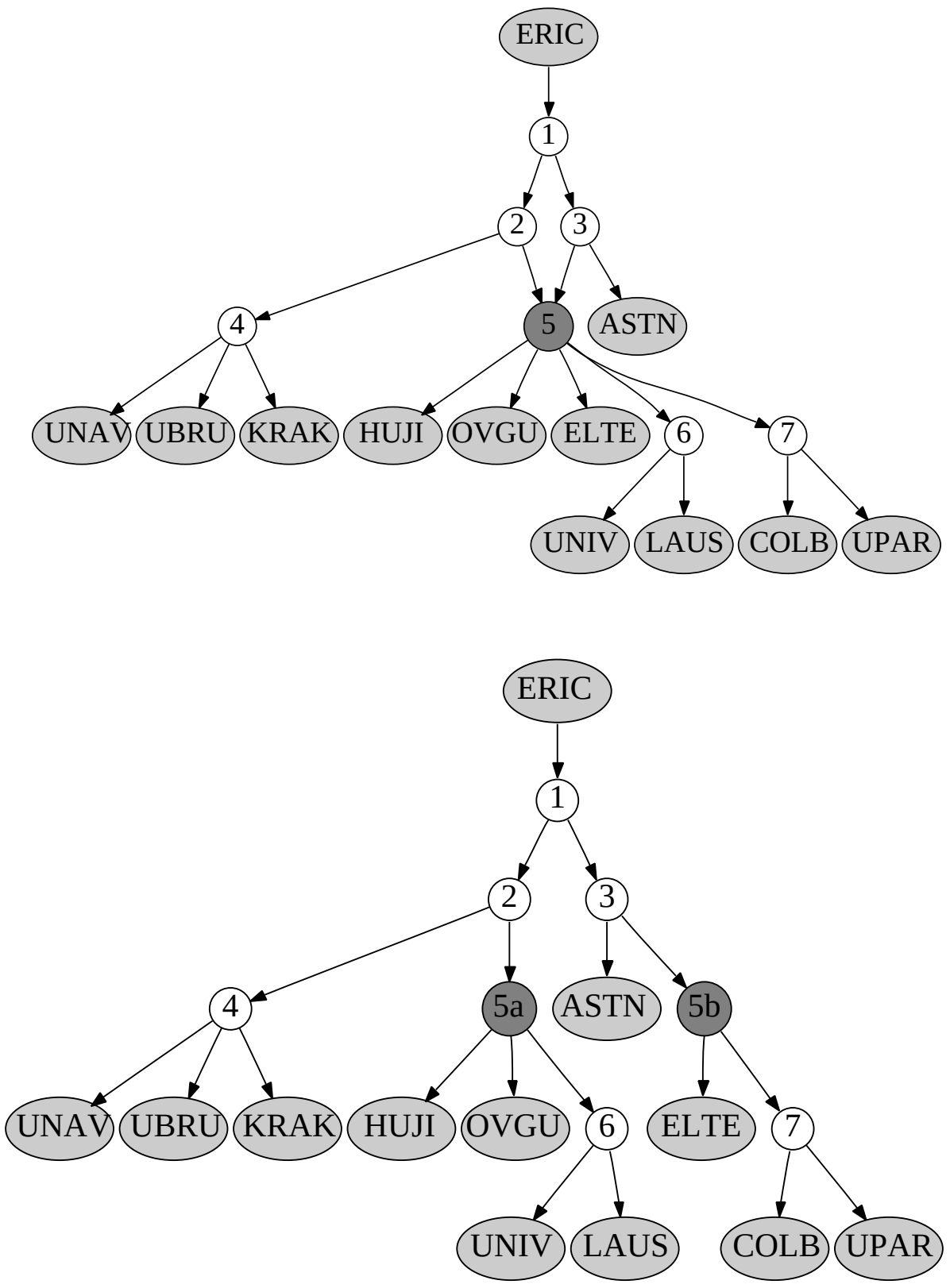

**Figure 6: Generating the proper tree graph. In the (upper graph) example shown probes sent from measurement host "ERIC" diverge at branching point 1. There are probes that are routed 1–2–5–6, and others as 1–3–5–7. Probes coming from nodes 2 and 3 will meet in the queue of node 5 with a negligeable probability since the difference of the propagation delays along the two different paths is much larger than the queuing delay in the queue of node 5. In this case we have to clone node 5 and introduce 5a and 5b as two separate nodes. The resulting ranching tree graph (lower graph) is then suitable for the delay tomography analysis.**

For delay tomography purposes the one-way delay of packets in batches of probes needs to be measured. In our measurement batches of back-to-back UDP packets are sent out addressed to the different destinations. Both IP addresses and port numbers are fixed. Packets are time stamped both at the sender and receiver hosts by the GPS synchronized DAG cards. The precision of stamps is about 100 ns at each side in absolute global time. The one-way delay is given by the difference of timestamps. Back-to-back packets are used in order to ensure that packets travel together on the shared part of their network paths, experiencing approximately the same network delay conditions. For this purpose and also to avoid self-congestion small size UDP packets (48 bytes) are used. By this method the number of probe packets sent in a batch will scale linearly with the number of destinations $N$ involved in the experiment. In our experiments $N = 13$ on the average. Measurements from the different sources are running simultaneously. 10000 probe batches are sent to the network of which less than 0.5 % are lost due to the asynchronous process start or finish at the measurement sites. Between consecutive batches 10 ms time gap is kept to let the network relax and ensure the independence criterion in Eq. (3). Experiments of this type are running regularly since July 2006 in the ETOMIC infrastructure several times a day. Since then a collection of data (collected from 850 experiments) is evaluated and stored in the ETOMIC database[15].

# 5. IMPLEMENTATION OF THE INFERENCE ALGORITHM

In this section we present the numerical implementation of the inference algorithm. Since we collect a large amount of raw data of end-to-end delay times between several machines, and we provide also the inferred probability distributions of the inner links online[7], the implementation should give a preview of the inferred inner link delay distributions available in a few seconds. If time and CPU resources allow, the delay distributions must be later refined at higher resolutions. We note here, that the link delay distribution inference is calculated independently for each sender host, hence the topology and inner link delay distributions can be calculated in parallel.

## 5.1 Successive approximation

A simple solution for both of the above demands is an iterative method, that applies the presented ML inference with different bin sizes. First a very raw estimation of the link delay distribution is calculated. For this first estimate we are using only a few large size bins and the ML calculation starts from the uniform distribution. Note that any other distribution can be used here that initializes each bin with nonzero values. After having a raw estimate of the distributions, a successive approximation phase begins. Here the bin size is reduced to the final resolution in successive steps according to a predefined series. For each bin size the ML algorithm starts from the smoothed version of the distributions calculated at the previous level and runs until the convergence criterion is satisfied. When the ML algorithm converges for the smallest prescribed bin size the calculation finishes.

This successive method has several advantages: i) The calculation can be interrupted at any bin size and later it can be refined again easily without storing and maintaining any other inner variables of the evaluation. ii) A further advantage is that the resulting distributions at each step of the approximation can be used for checking the convergence criterion of the ML estimation. iii) Network topology is fixed during the approximation steps. One can observe a given link and compare the delay distributions that were calculated at different bin sizes. From the difference between the distributions it is possible to estimate the error of the distribution parameters. Reusing the parameters of former distribution instead of using eg. uniform distribution as the initial distribution for the ML estimator will also speed up the inference process, see Table. 1.

Finally we mention, that it is possible to accelerate the overall convergence of the calculation by choosing the bin size sequence properly. The runtime required by the algorithm for a distribution with $B$ number of bins scales with $B^2$, since the calculation time of the ML estimation is mostly determined by the calculation of the convolution integrals

| Number of bins | Acceleration |
|---|---|
| 50 | 0.704 |
| 100 | 0.606 |
| 200 | 0.599 |
| 400 | 0.439 |

**Table 1: Acceleration gained in the runtime of our ML algorithm for cases with and without the knowledge of the previous, coarser time resolution at various number of bins. (Larger number of bins means higher resolution.)**

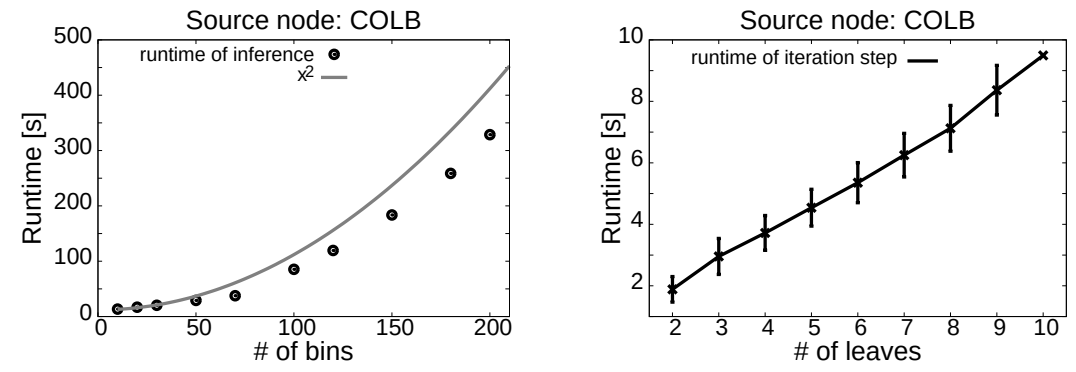

**Figure 7: Performance analysis of the ML implemented. Left: Varying the time resolution of the distributions ie. increasing the number of bins is investigated. The total runtime of the algorithm scales better than $\mathcal{O}(B^2)$. We show the results for the tree graph of the source node "COLB". Right: Scaling analysis. The runtime of the ML algorithm is tested against the number of receiver nodes in the network. Results for the sender node "COLB" are shown. The runtime increases approximately linearly with the number of leaves in the tree. The variance of the runtime denoted by the error bars grows even slower.**

Eq. (14). The convolutions have to be evaluated in each inner EM step of the ML algorithm. The execution time then scales with $sB^2$, where $s$ is the number of EM iteration steps at a given bin size. Choosing proper initial distribution with a given bin size $B_i$ results in fast convergence, which needs only a few ML steps $s_i$. For a given - usually exponentially decreasing - sequence of bin sizes (or $B_1, B_2, \ldots, B_k$ number of bins) the total runtime $s_1 B_1^2 + s_2 B_2^2 + \ldots + s_k B_k^2$ grows much slower than $sB_k^2$, where $s$ is the number of steps needed if we run the calculation immediately at the largest number of bins $B_k$. This is supported by the numerical evidence shown in Fig. 7.

From the point of view of large scale deployment of network tomography as a monitoring tool in the future, an important property is the scaling of the runtime of the inference algorithm with the number of monitoring nodes. The great advantage of the message-passing algorithms is that they have good scaling properties. It is expected that computational complexity grows only linearly with the longest path in the tree. In our experiment we studied the size dependence of the runtime of the inference algorithm. In measurement trees corresponding to various sender nodes we changed the number of receiver nodes from 2 to 10. In each case we repeated the analysis for all possible selection of nodes and averaged the resulting run-times. In Fig. 7 we show a typical result for one of the sender nodes. The observed linear growth of the runtime with the number of re-ceiver end nodes affirm the expected good scaling properties of the new method.

# 6. NEW SCALING PROPERTIES OF THE QUEUING DELAY

In the past decade it became obvious that Internet has several interesting fractal and scaling properties, such as the self-similar nature of traffic[18] and the power law scaling of the network topology[13]. The accelerated inference algorithm now makes it possible for us to get higher resolution distributions than it was possible in our earlier works[8, 9] and to find interesting new scaling properties in our data. We highlight that our distributions show signs of self-similar traffic on the links. Then we also establish a new scaling law for the distribution of the average link delays.

In Fig. 8 we show one of the tree graphs with the inferred queueing distributions. We can interpret some of the main distribution types of the Figure. For some of the links no distribution is shown. These are empty links inside the very high speed (≈10 Gbps) part of the GEANT network. The first link originating from the root node "ELTE" is a single up-link hop. The corresponding single hop distribution shown in Fig. 1 is a typical single queue delay distribution in case of self-similar input traffic. A Weibull distribution can be fitted and the Hurst exponent can be determined via the formula of Norros[14]

$$P(X > x) \sim \exp\left(-\frac{(C-m)^{2H}}{2k^2(H)c_v m^2}x^{2-2H}\right), \qquad (21)$$

where $C$ is the link capacity, $m$ the input traffic average rate, $c_v = \sigma/m$ the standard deviation divided by the input traffic mean and $k(H) = H^H(1-H)^{1-H}$, being $H$ the Hurst parameter. So far similar analysis was possible only on passively collected single router delay data[11, 10].

In Fig. 9 a typical distribution for a compound link consisting of many hops is shown. This and all the other downlink distributions are convolutions of the distributions of the hops in the compound link. The distribution is similar to a typical one-way end-to-end delay distribution. This distribution can also be well fitted by a Weibull distribution, though the Hurst exponents of the links have no simple relation to the Weibull parameter in this case.

It is possible now also to analyze the measured 183 different delay distributions produced in the measurements carried out from 13 sources. The dataset now covers the whole European continent and makes possible to draw some preliminary conclusions on the spatial distribution of queuing delays in a large network. Timescales of the distributions range over three orders of magnitude. A good overall representation of the data can be achieved by plotting the standard deviation of the distributions against their averages (see Fig. 10). We can see that the standard deviation scales linearly with the average over a wide timescale range. This means that the various distributions come approximately from the same distribution family with little variation in shape. This is consistent with our previous observation that the Weibull family fits well individual distributions, where standard deviation and average are related by

$$\sigma(x_i) = g(a)E(x_i), \qquad (22)$$

where $a$ is the shape parameter of the Weibull distribution and $g(a) = \frac{(\Gamma(1+2/a)-\Gamma^2(1+1/a))^{1/2}}{\Gamma(1+1/a)}$ is the prefactor. The fitted shape parameters for various distributions are in the $a = 0.5 - 2.0$ range thus the prefactor is in the $g = 0.5 - 1.1$ range in accordance with Fig. 10.

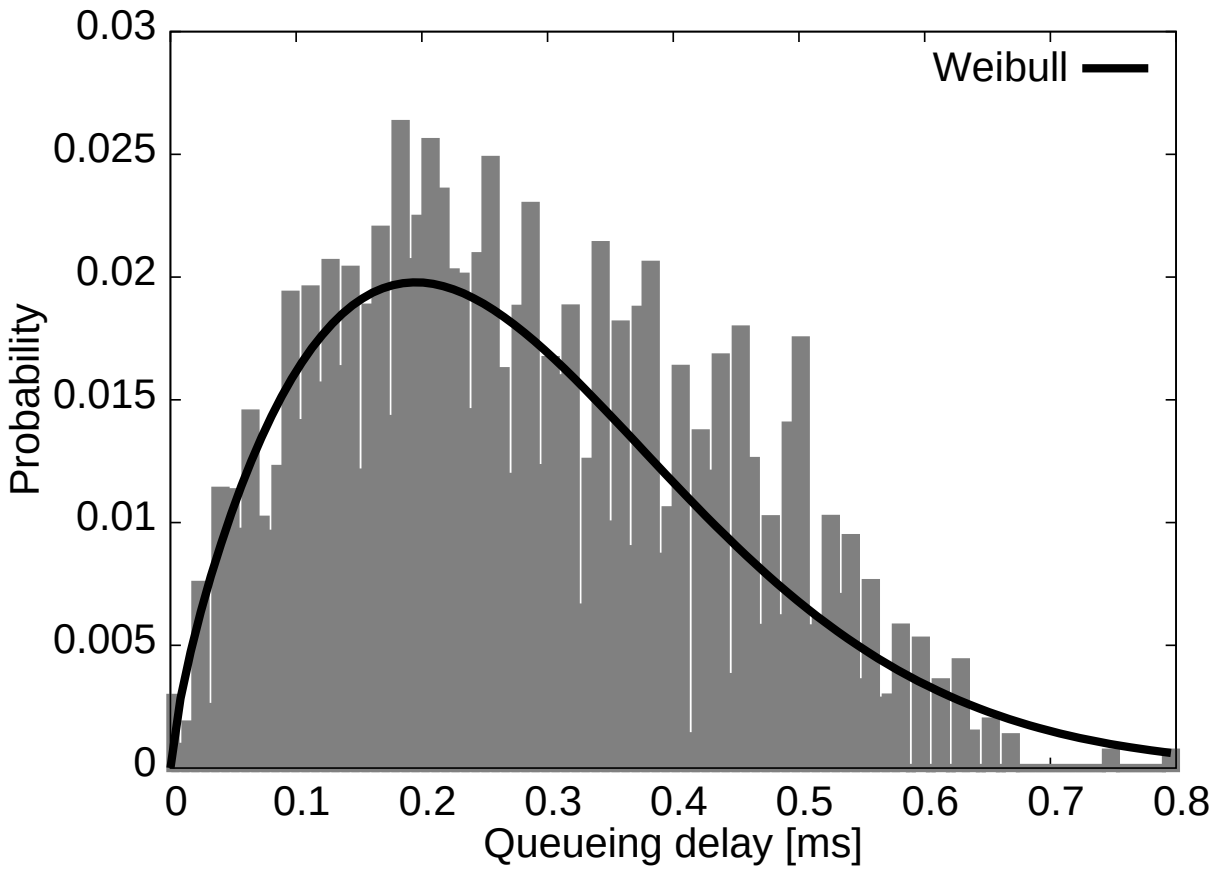

**Figure 9: Inferred queueing delay distribution for the compound link 2-ERIC. This is a typical downlink from the high speed core down to the 100 Mbps local ethernet of the measurement station "ERIC". A fitted Weibull function is also shown.**

Perhaps the most interesting result comes from the analysis of the distribution of average queuing delays. In Fig. 2 we show the complementary cumulative distribution of average delays measured on the 183 resolved links. On a semi-logarithmic scale it falls on a straight line indicating that

$$P(E(X) > x) \sim C_1 - C_2 \log(x), \qquad (23)$$

in the measurement range $x_{min} < x < x_{max}$. This means that the corresponding density shows $\sim 1/x$ type of power law scaling in the same temporal range. Such distribution is very special. Without upper and lower cutoffs it has no finite variance or mean and it is not even normalizable.

While the exact reason for this scaling is not yet obvious for us, a possible explanation can be that the new scaling reflects the scaling properties of link capacities built into the network. The queueing delay time on the links can be approximated with

$$x \approx \frac{P}{C-m}, \qquad (24)$$

which is the time a packet needs to pass trough a queue where the available bandwidth is $C - m$, where $C$ is the link capacity and $m$ is the input traffic average rate in the measured interval. For lightly loaded links it can be further approximated as $x \approx P/C$ and we can suspect that the $1/x$ scaling observed in the queueing delay is caused by the $1/C$ type scaling of the link capacities in a large scale network.

# 7. CONCLUSIONS

In this paper we introduced a new rigorous recursive algorithm for the computation of the likelihood and its numerical maximization through the EM algorithm in network tomography. We showed on standard numerical examples that its computational performance is better than that of previous methods. We implemented the algorithm and analyzed real high precision tomography measurements carried out in the ETOMIC infrastructure. It has been shown that the runtime of the algorithm scales with the system size. We managed to get high resolution queueing delay distributions from the European academic and research network GEANT. On single hop links the Hurst exponent of the self-similar input traffic has been determined from the exponent of the fitted Weibull distribution. Weibull distributions with various shape parameters give good approximation of the delay distributions of compound down-links. A new extreme scaling property of the investigated part of the Internet has been discovered in the averages of the delay distributions.

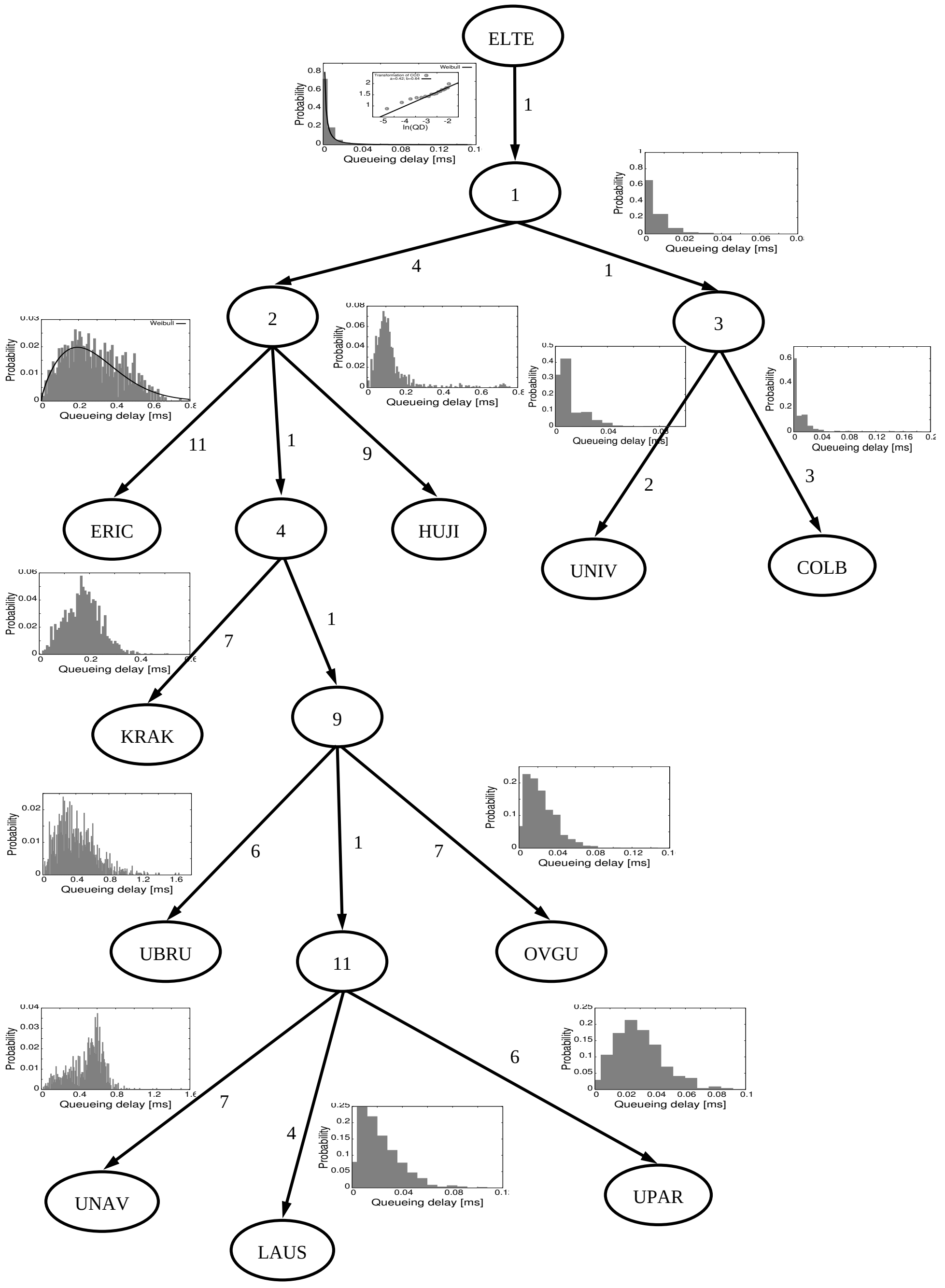

**Figure 8: A typical tree graph with the inferred queueing delay distributions. Only the nontrivial distributions are shown. Hop distance between neighboring routers are labelled. Branching nodes, anonymized as 1,2,3,4,9 and 11, are in the high speed (≈10 Gbps) part of the academic research network GEANT. The queuing delays between these routers are below the resolution of the measurement and are not shown. The ELTE-1 link is a single hop. We analyze its distribution in detail in Fig. 1. In Fig. 9 we investigate the compound link 2-ERIC, which is a typical multiple hop down-link.**

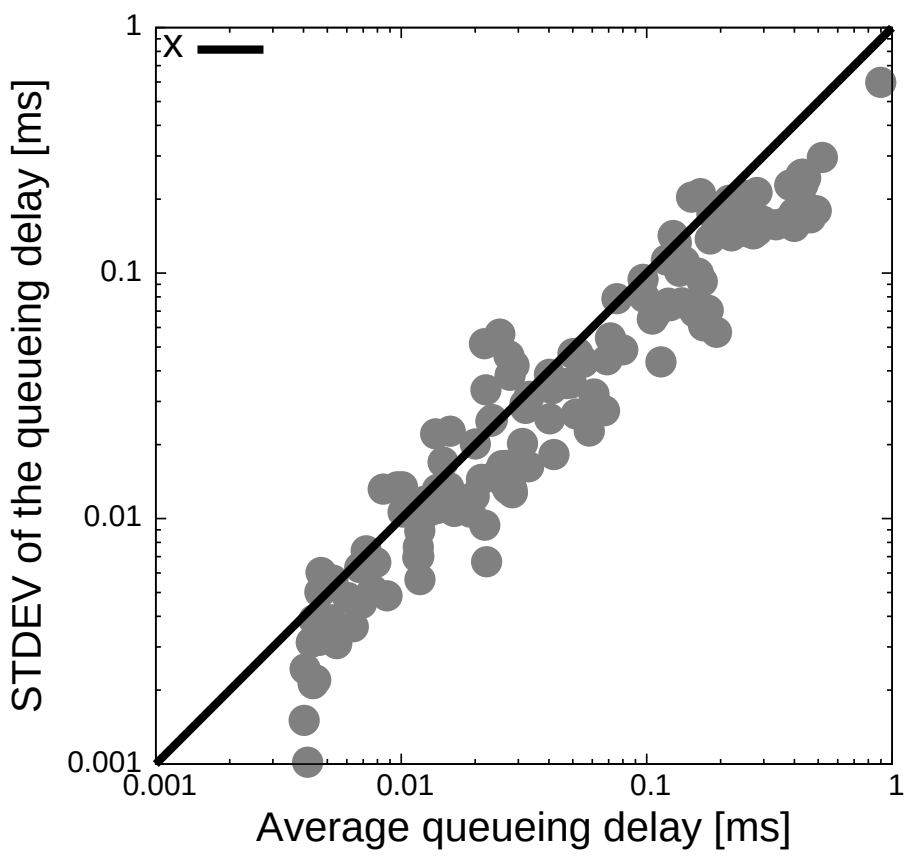

**Figure 10: The standard deviation as a function of the average of the queueing delay distributions. The straight line is $\sigma(x_i) = E(x_i)$. standard deviation values are in the $0.5E(X) < \sigma(X) < 1.1E(X)$ range, except for the smallest averages (coming from the high speed links), where the standard deviation is inaccurate due to the limited resolution of the measurement.**

The authors thank the partial support of the National Office for Research and Technology (NAP 2005/ KCKHA005) and the EU IST EVERGROW Integrated Project.

## 8. REFERENCES

[1] B. Augustin, X. Cuvellier, B. Orgogozo, F. Viger, T. Friedman, M. Latapy, C. Magnien and R. Teixeira. Avoiding traceroute anomalies with paris traceroute. In *Proceedings of the 6th ACM SIGCOMM on Internet measurement (Rio de Janeriro, Brazil)*, pages 153–158.

[2] R. Caceres, N. Duffield, J. Horowitz, and D. Towsley. Multicast-based inference of network-internal loss characteristics. In *IEEE Transactions on Information Theory*, volume 45, 1999.

[3] M. Coates and R. Nowak. Network tomography for internal delay estimation. In *In Proceedings of IEEE International Conference on Acoustics, Speech and Signal Processing, (Salt Lake City, UT, May 2001.)*, 2001.

[4] D. Morato, E. Magana, M. Izal, J. Aracil, F. Naranjo, F. Astiz, U. Alonso, I. Csabai, P. Haga, G. Simon, J. Steger and G. Vattay. Etomic: A testbed for universal active and passive measurements. In *Proceedings of TRIDENTCOM (23-25 February 2005, Trento, Italy 2005; Best Testbed Award)*, pages 283–289.

[5] B. J. Frey. *Graphical models for machine learning and digital communication.* The MIT Press, Cambridge, Mass, 1998.

[6] *http://gaia.cs.umass.edu/minc*

[7] *http://www.etomic.org*

[8] P. Haga, I. Csabai, G. Simon, J. Steger and G. Vattay. Measuring the dynamical state of the internet: Large scale network tomography via the etomic infrastructure. *Complexus*, 3:119, 2006.

[9] P. Matray, G. Simon, J. Steger, I. Csabai, P. Haga and G. Vattay. Results of large-scale queueing delay tomography performed in the etomic infrastructure. In *Proceedings of the 9th IEEE Global Internet Symposium*, 2006.

[10] M. Izal and J. Aracil. On the influence of self-similarity on optical burst switching traffic. In *Proceedings of GLOBECOM*, volume 3, pages 2308–2312.

[11] K. Papagiannaki, S. Moon, C. Fraleigh, P. Thiran, F. Tobagi and C. Diot Analysis of measured single-hop delay from an operational backbone network. In *Proceedings of INFOCOM*, 2002.

[12] G. Liang and B. Yu. Maximum pseudo-likelihood estimation in network tomography. In *IEEE Transactions on Signal Processing, 51(8):2003.*, 2003.

[13] M. Faloutsos, P. Faloutsos and C. Faloutsos. On power-law relationships of the internet topology. In *Proceedings of ACM SIGCOMM, (Boston)*, 1999.

[14] I. Norros. On the use of fractional brownian motion in the theory of connectionless networks. *IEEE Journal on Selected Areas in Communications*, 16(3):953–962, August 1995.

[15] P. Matray, I. Csabai, P. Haga, J. Steger, L. Dobos and G. Vattay Building a prototype for network measurement virtual observatory. In *Joint International Conference on Measurement and Modeling of Computer Systems archive; Proceedings of the 3rd annual ACM workshop on Mining network data (San Diego, California)*, pages 23–28.

[16] F. Presti and N. Duffield. Multicast-based inference of network-internal delay distributions. In *ACM/IEEE Transactions on Networking*, 2002.

[17] Y. Vardi. Network tomography: estimating source-destination traffic intensities from link data. *Journal of the American Statistical Association*, 91:365, 1996.

[18] W. E. Leland, M. S. Taqqu, W. Willinger and D. V. Wilson On the self-similar nature of ethernet traffic. In *IEEE/ACM Transactions on Networking*, pages 1–15.

This document is an example of BibTeX using in bibliography management. Three items are cited: *The LaTeX Companion* book [1], the Einstein journal paper [2, 3, 4, 5, 6].

# References

[1] Michel Goossens, Frank Mittelbach, and Alexander Samarin. *The LaTeX Companion*. Addison-Wesley, Reading, Massachusetts, 1993.

[2] Gabor Vattay, Dennis Salahub, Istvan Csabai, Ali Nassimi, and Stuart Kauffman. Quantum criticality at the origin of life. *Journal of Physics*, 626, 2015.

[3] Péter Mátray, Péter Hága, Sándor Laki, Gábor Vattay, and István Csabai. On the spatial properties of internet routes. *Computer Networks*, 56(9):2237–2248, 2012.

[4] Sarika Jalan, Norbert Solymosi, Gábor Vattay, and Baowen Li. Random matrix analysis of localization properties of gene coexpression network. *Physical Review E*, 81(4):046118, 2010.

[5] Peter Matray, Istvan Csabai, Peter Haga, Jozsef Steger, Laszlo Dobos, and Gabor Vattay. Building a prototype for network measurement virtual observatory. In *Proceedings of the 3rd annual ACM workshop on Mining network data*, pages 23–28. ACM, 2007.

[6] Predrag Cvitanović, Niels Søndergaard, Gergely Palla, Gábor Vattay, and CP Dettmann. Spectrum of stochastic evolution operators: Local matrix representation approach. *Physical Review E*, 60(4):3936, 1999.